\documentclass[final,5p,twocolumn,10pt,authoryear]{elsarticle}

\usepackage{booktabs}
\usepackage{url}

\usepackage{breakurl}
\usepackage[breaklinks]{hyperref}
\usepackage[colorinlistoftodos,prependcaption,textsize=footnotesize]{todonotes}

\usepackage[utf8x]{inputenc}
\usepackage[russian,english]{babel}
\usepackage{multirow}
\usepackage{dingbat}
\usepackage{tabularx}
\usepackage{amssymb}
\usepackage{pifont}
\usepackage{natbib} 
\usepackage{hyperref}
\hypersetup{colorlinks,linkcolor={blue},citecolor={blue},urlcolor={blue}}  

\usepackage{footnote}
\makesavenoteenv{tabular}
\makesavenoteenv{table}
\def\BibTeX{{\rm B\kern-.05em{\sc i\kern-.025em b}\kern-.08emT\kern-.1667em\lower.7ex\hbox{E}\kern-.125emX}}
    
\providecommand{\keywords}[1]
{
  \small	
  \textbf{\textit{Keywords---}} #1
}

\begin{document}

%
\title{From Seaweed to Security: The Emergence of Alginate in Compromising IoT Fingerprint Sensors}

%


\author[add1]{Pouria Rad}
\ead{peslamirad@augusta.edu}


\author[add2]{Gokila Dorai}
\ead{gdorai@augusta.edu}


\author[add2]{Mohsen Jozani}
\ead{mjozani@augusta.edu}

\address[add1]{School of Computer and Cyber Sciences, Augusta University, GA-30901, USA}

%
\begin{abstract}
The increasing integration of capacitive fingerprint recognition sensors in IoT devices presents new challenges in digital forensics, particularly in the context of advanced fingerprint spoofing. Previous research has highlighted the effectiveness of materials such as latex and silicone in deceiving biometric systems. In this study, we introduce Alginate, a biopolymer derived from brown seaweed, as a novel material with the potential for spoofing IoT-specific capacitive fingerprint sensors. Our research uses Alginate and cutting-edge image recognition techniques to unveil a nuanced IoT vulnerability that raises significant security and privacy concerns. Our proof-of-concept experiments employed authentic fingerprint molds to create Alginate replicas, which exhibited remarkable visual and tactile similarities to real fingerprints. The conductivity and resistivity properties of Alginate, closely resembling human skin, make it a subject of interest in the digital forensics field, especially regarding its ability to spoof IoT device sensors. This study calls upon the digital forensics community to develop advanced anti-spoofing strategies to protect the evolving IoT infrastructure against such sophisticated threats.

%

\end{abstract}
\maketitle

\section{Introduction} \label{intro}
The Internet of Things (IoT) represents a significant evolution in interconnected, intelligent devices, revolutionizing industries from healthcare to transportation. However, ensuring their security becomes paramount as these devices become more integrated into our daily lives. Biometric authentication, particularly fingerprint recognition, is a standard security method in IoT devices. Fingerprint recognition, especially capacitive fingerprint recognition, offers a unique and personalized method of authentication, leveraging the distinct patterns found on every individual's fingertips. Yet, studies by \cite{lee2017fingerprint, bowden2012fooling} and \cite{hossain2016biometrics} highlight the the vulnerabilities of such authentication systems against fingerprint spoofing and at the same time their wide usage for biometric-based security solutions in IoT systems.

Historically, various materials have been explored for fingerprint spoofing attacks, including latex, silicone, and gelatin, however, a new competitor has emerged. Alginate, a naturally occurring biopolymer derived from seaweed, has gained prominence in bio-printing or skin mimicry due to its unique properties. It closely mimics the texture and elasticity of human skin, making it an ideal material for creating fingerprint replicas \citep{xu2020advances}. Moreover, its wide availability and ease of use further accentuate its potential as a spoofing material. Studies have shown that Alginate-based hydrogels present properties suitable for bioprinting of skin models \citep{zidarič2020polysaccharide}, and its combination with other materials can produce skin-like hydrogels with high elasticity and sensitivity \citep{cavallo2023marine}. Using Alginate molds, we can cast fingerprints to form a skin-like cast that needs no further post-process to be recognized by the fingerprint sensors. This amplifies a vulnerability challenge for security of fingerprint sensors which needs to be addressed.

While these materials have been used to deceive biometric systems of even advanced authentication mechanisms, the digital age has further complicated this scenario. The study by \cite{jozani2022biometric} explored the Etsy Platform, where users often share high-resolution images of their hands, which can inadvertently become sources for biometric data extraction using AI and machine learning tools. In their study, they highlighted the risks of posting biometric information online and the lack of adequate protective measures by social media and marketplace companies. With advanced image processing, these fingerprints can be extracted from such photographs, presenting a substantial security concern \citep{marasco2015survey}, permanently compromising biometric security for social media users. This exposure is particularly concerning because, unlike passwords, compromised biometric data, such as fingerprints, remains permanently vulnerable \citep{ratha2001enhancing, aounallah2022privacy}. Past breaches highlight this enduring risk \citep{cavoukian2012biometric} and the combination of high-definition camera technology and AI-driven tools magnifies this vulnerability, allowing automated extraction from vast datasets \citep{engelmann2018media}. This underscores the pressing need to address these new age vulnerabilities, especially within forensics contexts.

Therefore, to guide our exploration, we pose the following research question: "How effective is Alginate in spoofing capacitive fingerprint sensors in IoT devices, and what implications does this have for the security of biometric data and the integrity of digital forensic investigations?" This inquiry not only directs our experimental approach but also frames our discussion on the evolving landscape of IoT security and digital forensic practices.

In this study, we explore the efficacy of Alginate in fingerprint spoofing. While our experiments primarily utilize consensually obtained real-life samples of fingerprint molds, the broader implications of our findings resonate with the digital forensics community. The potential for using publicly available images for fingerprint data extraction to fabricate fingerprint replicas using easily accessible technologies like 3D printing introduces novel challenges that must be addressed to maintain the integrity of forensic investigations and biometric security, particularly in IoT devices. Our research underscores the need for forensic experts to adapt to these emerging threats by understanding the nuances of this vulnerability, pushing stakeholders in the IoT industry to develop more robust security measures, ensuring that devices remain secure even in the face of sophisticated attacks.

The remainder of this paper is structured as follows. Section~\ref{sec:related} provides a detailed review of the related work, highlighting the findings of previous studies and identifying gaps. Section~\ref{sec:method} describes the methodology employed in our research, detailing the experiments conducted and the rationale behind them. Section~\ref{sec:findings} presents our findings, followed by a discussion in Section~\ref{sec:discussion}. Limitations of this work are presented in Section~\ref{sec:limitations}. Finally, Section~\ref{sec:conclusion} concludes the paper, offering recommendations for future research and potential solutions to the identified vulnerability.

\begin{table*}[h]
\caption{Summary of Related Works}
\centering
\renewcommand{\arraystretch}{1.3} 
\resizebox{\textwidth}{!}{
\begin{tabular}{|p{0.15\textwidth}|p{0.8\textwidth}|p{0.3\textwidth}|}
\hline
\textbf{Aspect} & \textbf{Description} & \textbf{Reference} \\
\hline
\multicolumn{3}{|c|}{\textbf{Biometrics Aspects}} \\
\hline
Uniqueness & Utilizes individual patterns in fingerprints for secure authentication methods. & \cite{hossain2016biometrics} \\
Integration & Non-intrusive integration into smartphones, smart locks, etc. & \cite{lee2017fingerprint} \\
Challenges & Faces standardization challenges in IoT; requires active user interaction, poses data breach risks. & \cite{shahriar2021lightweight} \\
Vulnerabilities & Includes risks from partial systems and smudges on screens that can reconstruct fingerprints. & \cite{roy2017masterprint}, \cite{lee2017fingerprint} \\
Solutions & Introduction of lightweight fuzzy extractors and multimodal authentication for better security. & \cite{shahriar2021lightweight}, \cite{macek2016multimodal} \\
Data Protection & Methods to ensure biometric data protection during transmission and storage. & \cite{golec2022biosec} \\
\hline
\multicolumn{3}{|c|}{\textbf{Material/Method}} \\
\hline
Materials & Describes materials like latex, silicone, gelatin, and alginate for creating replicas. & \cite{matsumoto2002impact}, \cite{galbally2010evaluation}, \cite{cappelli2007fake}, \cite{xu2020advances} \\
3D Printing & Advanced techniques for creating detailed fingerprint replicas; also high-resolution image threats. & \cite{neubauer2016survey}, \cite{jozani2022biometric} \\
\hline
\multicolumn{3}{|c|}{\textbf{Vulnerability Aspects}} \\
\hline
Spoofing & Discusses sensors being deceived by material replicas; challenges in liveness detection. & \cite{marasco2015survey}, \cite{ghiani2013fingerprint} \\
Residual Fingerprints & Risks from smudges left on sensors; high-resolution sensors increase sensitivity to spoofing. & \cite{lee2017fingerprint}, \cite{antonelli2015fake} \\
IoT Integration & Highlights the resource constraints and security standardization challenges in IoT. & \cite{hossain2016biometrics} \\
\hline
\multicolumn{3}{|c|}{\textbf{Concerns}} \\
\hline
Digital Exposure & Risks from high-resolution images on social platforms; digital forensics and machine learning techniques extracting fingerprint details. & \cite{jozani2022biometric}, \cite{marasco2015survey}, \cite{ross2016handbook}, \cite{engelmann2018media} \\
Preventative Measures & Emphasizes the need for stricter privacy policies and user education on sharing biometric data. & \cite{ratha2001enhancing} \\
Data Compromise & Discusses the lasting threat of biometric data breaches and database vulnerabilities. & \cite{ratha2001enhancing}, \cite{cavoukian2012biometric} \\
Forensics Solutions & Advocates for biometric template protection and fusion of multiple biometrics to enhance security. & \cite{patel2015secure}, \cite{ross2003information} \\
\hline
\end{tabular}}
\label{tab:related_works}
\end{table*}

\section{Related Work} 
\label{sec:related}

The domain of biometric authentication in IoT has garnered significant attention due to its potential to offer secure and user-friendly mechanisms for device access. This section provides a comprehensive review of the existing literature, focusing on the evolution, strengths, vulnerabilities, and potential threats associated with biometric authentication in IoT. We have summarized related works in Table~\ref{tab:related_works}.

The IoT ecosystem is characterized by a vast array of devices, each with its unique functionalities and purposes. As these devices become more integrated into our daily lives, the need for secure and reliable authentication mechanisms becomes paramount. Among the various authentication methods, biometric authentication, particularly fingerprint recognition, has emerged as a preferred choice for many IoT devices.

Fingerprint recognition offers several advantages. Firstly, it leverages the uniqueness of an individual's fingerprint patterns, ensuring that the authentication process is personalized and secure \citep{hossain2016biometrics}. This uniqueness is a cornerstone of biometric security, making it less vulnerable to breaches in IoT systems or infrastructure. Secondly, fingerprint sensors are compact and can be seamlessly integrated into a wide range of devices, from smartphones to smart door locks. This integration ensures that the user experience remains intuitive and non-intrusive \citep{lee2017fingerprint}.

However, the adoption of fingerprint sensors in IoT devices is not without challenges. The diverse range of IoT devices, each with its computational and sensing capabilities, poses a challenge in standardizing biometric solutions. Moreover, while fingerprints offer stability, they demand active user interaction, which might not always be feasible for continuous authentication. Additionally, the risk of fingerprint data being compromised, either through physical means or digital extraction, remains a significant concern \citep{shahriar2021lightweight}.

Several vulnerabilities have been identified in fingerprint-based authentication systems. For instance, partial fingerprint-based authentication systems have been found to be potentially vulnerable, especially when multiple impressions are enrolled per finger \citep{roy2017masterprint}. Furthermore, the risk posed by smudges left on device touch screens, which can be used to reconstruct an image of the enrolled fingerprint, has been highlighted in the literature by \cite{lee2017fingerprint}.

To address these challenges and vulnerabilities, researchers have been exploring innovative solutions. The introduction of lightweight fuzzy extractors, which offer a more error-tolerant biometric solution suitable for resource-constrained IoT devices, is one such advancement \citep{shahriar2021lightweight}. Additionally, the concept of multimodal biometric authentication, which combines multiple biometric traits, has been proposed to enhance the reliability and security of authentication systems \citep{macek2016multimodal}. With the increasing concerns about the security of biometric data, new methods have been introduced to ensure the protection of biometric data during transmission and storage \citep{golec2022biosec}. As the IoT landscape continues to evolve, so does the need for adaptable and resilient biometric solutions. The dynamic nature of this ecosystem necessitates continuous research and innovation. As devices become more sophisticated and threats evolve, the focus will inevitably shift towards developing more advanced and secure fingerprint authentication mechanisms.

While the advancements in biometric authentication for IoT devices, particularly fingerprint recognition, have brought about enhanced security, they are not without vulnerabilities. The very uniqueness of fingerprints, which makes them a reliable authentication method, also makes them a prime target for attackers. If an attacker can replicate a fingerprint with high fidelity, they can potentially gain unauthorized access to a device or system. Over the years, various materials and methods have been employed to spoof fingerprints, each with its own set of challenges and successes. As we delve deeper into the realm of fingerprint spoofing, it becomes imperative to understand the materials traditionally used, their effectiveness, and the innovative methods attackers might employ.

Fingerprint spoofing is the art and science of deceiving a biometric system by presenting a fabricated fingerprint that matches a legitimate user's fingerprint. The success of such an attack is contingent upon the quality and accuracy of the fabricated fingerprint, which is largely determined by the materials and methods employed for its creation.

Historically, a variety of materials have been employed to create fake fingerprints. Latex, for instance, is a flexible material that can be molded easily to replicate the fine ridges and valleys of a fingerprint. Its elasticity allows it to adapt to the pressure applied during the authentication process, making it a popular choice for spoofing attacks \citep{matsumoto2002impact}. Silicone, known for its durability and malleability, can be used to create highly detailed fingerprint replicas. Its texture closely resembles human skin, making it harder for sensors to distinguish between a real finger and a silicone spoof \citep{galbally2010evaluation}. Gelatin, often used in theatrical makeup, can be molded to create lifelike fingerprint replicas. Its soft texture and moisture content make it a suitable material for spoofing attacks, especially against sensors that rely on moisture detection \citep{cappelli2007fake}.

The evolution of fingerprint sensors and their increasing sophistication have necessitated the exploration of new materials and methods. Alginate, a biopolymer derived from seaweed, has emerged as a promising material in this context. Its properties, such as its texture and elasticity that closely mimic human skin, make it a promising material for fingerprint spoofing \citep{xu2020advances}. Its ability to capture fine details and its inherent conductivity further enhance its potential as a spoofing material.

In terms of methods, the process of fingerprint spoofing typically involves capturing a latent fingerprint, creating a mold, and then using the chosen material to cast a fake fingerprint. Advanced techniques, such as 3D printing, have been explored to create more accurate and detailed fingerprint replicas \citep{neubauer2016survey}. The digital age has further complicated the landscape. Platforms where users often share high-resolution images can inadvertently become sources for biometric data extraction, leading to potential spoofing attacks \citep{jozani2022biometric}.

It's worth noting that while the materials and methods for fingerprint spoofing have evolved, so have the countermeasures. Modern fingerprint sensors incorporate a range of anti-spoofing techniques, from analyzing the electrical properties of the skin to detecting blood flow and moisture content. However, as spoofing methods become more sophisticated, the challenge lies in staying a step ahead of potential attackers.

While understanding the materials and methods used in fingerprint spoofing is crucial, it's equally important to delve into the specific vulnerabilities of capacitive fingerprint recognition systems. These systems, commonly found in IoT devices, present their own set of challenges and vulnerabilities.
Capacitive fingerprint recognition systems operate by gauging the capacitance differences between the ridges and valleys of a fingerprint \citep{lee2017fingerprint}. The ridges, being proximate to the sensor, exhibit a different capacitance than the valleys, and this differential is employed to generate a unique fingerprint pattern. While these sensors are celebrated for their precision and resilience against external environmental factors, they are not impervious to vulnerabilities.

As mentioned before the primary vulnerability of capacitive fingerprint sensors is their susceptibility to spoofing attacks. These vulnerabilities accentuates the importance of integrating supplementary security measures, such as liveness detection. Liveness detection, which ascertains if the presented fingerprint originates from a living individual, has been proposed as a countermeasure against spoofing attacks. However, advanced spoofing materials and techniques can sometimes elude even these liveness detection mechanisms. As explained before, some materials can mimic the sweat pores and intricate details of a real finger, making it a challenge for liveness detection algorithms to distinguish between genuine and fake fingerprints \citep{ghiani2013fingerprint}.

Furthermore, the escalating resolution of fingerprint sensors, although enhancing accuracy, can also introduce vulnerabilities. High-resolution sensors can capture intricate details of a fingerprint, but they also become more sensitive to minute variations, which can be exploited by adversaries. For instance, an attacker might craft a spoof with exaggerated minute details, potentially increasing the likelihood of a successful attack \citep{antonelli2015fake}.

On the other hand, the integration of fingerprint sensors in IoT devices introduces additional challenges. Given the resource constraints of many IoT devices, implementing robust security measures becomes a challenge. Additionally, the diverse range of IoT devices, each with its unique hardware and software configurations, makes standardizing security solutions difficult. This diversity can lead to inconsistencies in security implementations, providing potential loopholes for attackers \citep{hossain2016biometrics}. While capacitive fingerprint recognition systems offer a reliable and efficient means of biometric authentication, they come with their set of vulnerabilities. Biometric authentication issues and vulnerabilities will evolve with the IoT and to secure and maintain biometric systems, researchers and industry professionals must keep current and innovate.

\section{Methodology} 
\label{sec:method}
The widespread integration of capacitive fingerprint identification technology in IoT devices has presented significant security challenges for the digital forensics sector. The objective of our study is twofold: (1) investigate the potential of Alginate in deceiving fingerprint sensors that are often seen in IoT devices; and, (2) explore the feasibility of extracting reproducible fingerprint data from digital photographs available on the internet. Our research delves into two primary areas: firstly, the effectiveness of Alginate in creating deceptive fingerprint impressions, and secondly, the innovative methodology of processing high-resolution photographs to replicate fingerprint data. This approach involves transforming detailed fingerprint patterns extracted from digital images into three-dimensional models, which then serve as molds for Alginate-based fingerprint replicas. By focusing on these specific areas, our study aims to highlight potential vulnerabilities in biometric authentication systems, particularly in IoT devices.

\subsection{Material Selection and Preparation}
The success of fingerprint spoofing relies heavily on the quality of the material used. The ability of a substance to mimic the tactile and electrical features of human skin may be a determining factor in its ability to fool fingerprint sensors. Many potential candidates were considered in the process of identifying suitable materials, each with their own set of advantages and disadvantages; however, Alginate stood out as particularly promising.

The unique qualities of Alginate were largely responsible for its selection. When mixed with water, it changes into a gelatinous material having remarkable similarities to the feel and pliability of human skin. Alginate's intrinsic conductivity and resistivity qualities are very similar to those of human skin, extending the similarities beyond superficial features. Therefore, Alginate presents itself as an excellent option for our experimental efforts.

\begin{figure}[t!]
  \centering
  \includegraphics[width=0.3\textwidth]{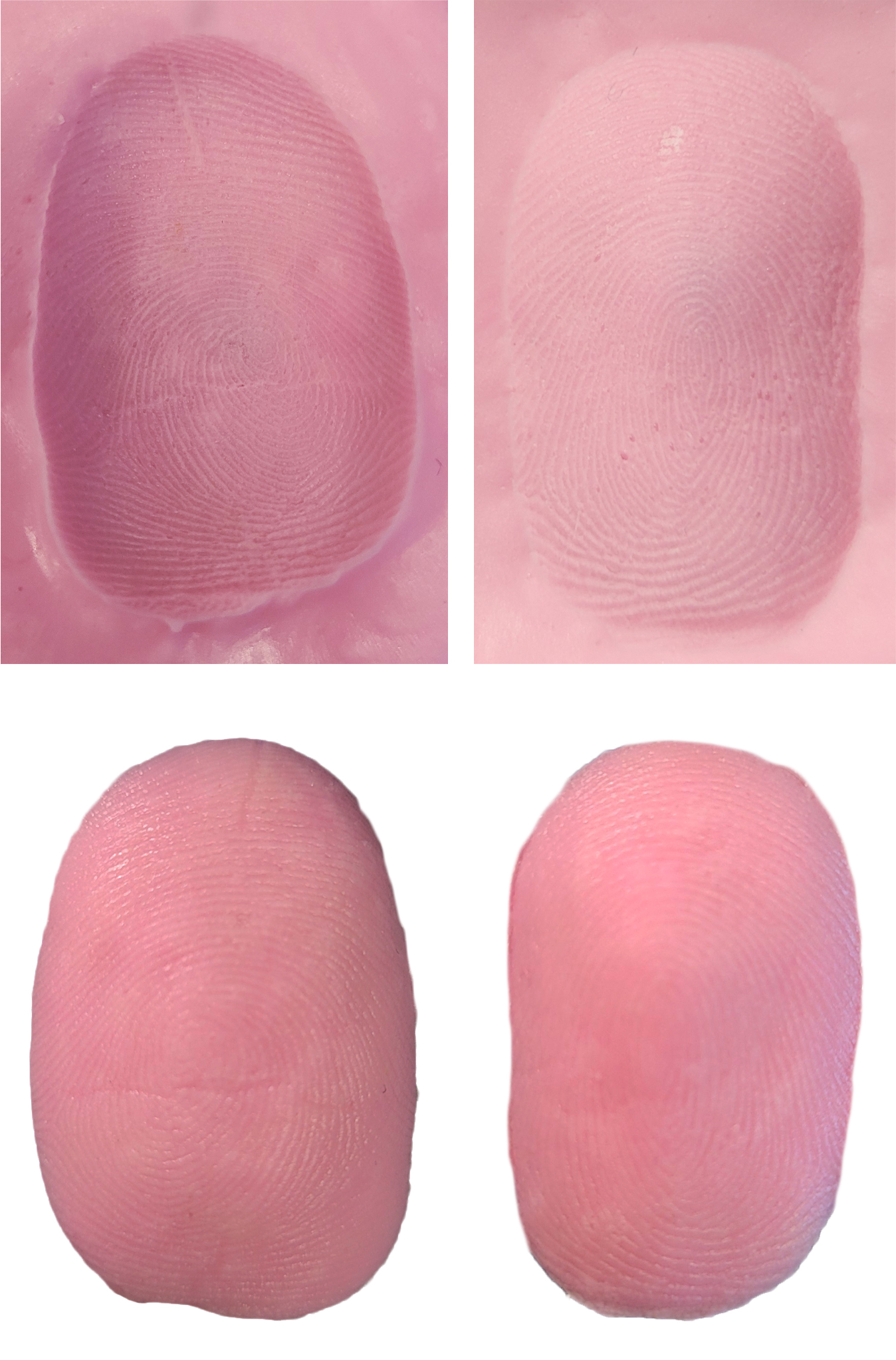}
  \caption{Example of fingerprint molds and casts made with Alginate}
  \label{figure1}
\end{figure}

In this phase of our set of experiments, we aimed to confirm Alginate's qualities in regard to fingerprint spoofing. To achieve this, we used real fingerprints collected from our lab members and staff to create castings. To get the desired consistency, Alginate powder was mixed with water according to a formula proposed by it manufacturer. After extensive agitation to ensure the mixture was smooth and free of clumps, it was poured into molds to capture the intricate patterns that are signatures of genuine fingerprints. The Alginate cast was carefully removed after it had been prepped for use in testing fingerprint sensors in Internet-of-Things gadgets. One of the sets of Alginate molds and casts used in our studies is shown in Figure~\ref{figure1}.

The use of Alginate was also substantiated due to its extensive accessibility and straightforward application. The non-toxic properties of the substance guarantee safety during its manipulation, while its rapid curing time expedites the testing procedure. In addition, the biodegradable properties of Alginate are in line with the increasing focus on sustainable research methodologies.

Our first investigations focused on the use of authentic fingerprint molds, but, we acknowledge the wider ramifications and potential risks associated with our study. In the following sections, we will explore the proposed idea of extracting fingerprint patterns from photos, transforming them into molds using 3D printing, and utilizing Alginate for casting. This analysis will emphasize the various vulnerabilities that may arise from this process.

\subsection{Fingerprint Extraction and 3D Printing}
In the contemporary era of widespread digital dissemination, it is possible for even an informal photograph to disclose biometric information unintentionally. This phenomenon is particularly true for visual representations that depict human hands or fingers since they might be susceptible to exploitation for the purpose of extracting fingerprint patterns. The procedure involved in collecting fingerprints from photographs and subsequently reproducing them through 3D printing is a complex and intricate operation.

The theoretical feasibility of utilizing publicly available images for fingerprint spoofing presents a unique and speculative challenge. One potential scenario involves the use of high-resolution images commonly shared on social media platforms, personal blogs, or public forums. These images, particularly those where individuals' hands are prominently displayed, might inadvertently reveal fingerprint details. The effectiveness of these images in spoofing attacks hinges on the camera's resolution and the clarity with which fingerprints are captured. Modern digital cameras and smartphones often possess the capability to take photographs with sufficient detail to potentially disclose fingerprint patterns.

Another significant consideration is the abundance of high-quality images of public figures or individuals frequently captured by high-definition media. These images, stemming from interviews, events, or photo shoots, could serve as potential sources for detailed fingerprint data due to the professional quality of media coverage. Moreover, in forensic investigations or journalistic contexts, where high-resolution images are analyzed for detailed information, there is a risk of such images becoming public, further exacerbating the potential for unauthorized fingerprint data extraction.

For fingerprint extraction, algorithms are the backbone. The YOLO (You Only Look Once) algorithm, known for its efficiency in object detection, was utilized in an earlier study by \cite{jozani2022biometric} to swiftly identify and crop hands and fingers from images on platforms like Etsy. However, the extraction is just the beginning. The quality of the fingerprint image often needs enhancement. Techniques such as histogram equalization and Fourier transform have been proposed to improve the clarity and contrast of these images \cite{gonzalez}. Another study highlighted the use of deep learning for fingerprint minutiae extraction, demonstrating the potential of neural networks in this domain \cite{darlow}.

Once the fingerprint images are enhanced, the next step is replication. 3D printing offers a solution. While FDM (Fused Deposition Modeling) printers build objects layer by layer, they might not offer the precision required for fingerprints. DLP (Digital Light Processing) printers, on the other hand, use light to solidify a liquid resin, providing a higher resolution suitable for our needs \cite{gibson}.

With the enhanced fingerprint images and a DLP 3D printer, we proposed creating molds that could emulate the fingerprint patterns. The ultimate goal was to determine if these 3D-printed molds, when used with Alginate, could deceive fingerprint sensors.

\subsection{IoT Device Selection}
Our investigation was limited to fingerprint-sensing house locks and padlocks, sometimes known as ``smart locks'' among IoT gadgets and devices. The prevalence of these devices is on the rise, providing homeowners with a combination of convenience and a sense of enhanced protection. The combination of keyless entry, apps, and customizable access restrictions, together with the added security provided by fingerprint verification, has contributed to the widespread adoption of these devices.

However, as with many technological advancements, there's a spectrum of quality and security. Not all fingerprint-enabled home locks are created equal. Research indicates that the affordability of IoT devices often correlates with their vulnerability to security attacks. This is primarily due to the use of lower-quality fingerprint sensors in less expensive models. For instance, a study by \cite{kumar2019quality} revealed that more economical biometric devices tend to compromise on sensor quality, rendering them more susceptible to spoofing attacks. Additionally, their study highlighted that budget constraints in manufacturing low-end IoT devices often lead to a trade-off where functionality is prioritized over security, thus creating exploitable vulnerabilities.

Given this backdrop, our selection criteria for IoT devices was twofold. First, we aimed to choose devices that are commonly available and widely used by consumers. This ensures that our findings have broad relevance. Second, we sought a mix of devices across different price points, allowing us to better understand the impact of our proposed vulnerability. Our methodology's objective was to provide a comprehensive understanding of the vulnerabilities present in commonly used IoT home locks. By focusing on these devices, we hoped to shed light on potential security risks that millions of homeowners might unknowingly be exposed to.

\subsection{Attack Implementation}
In our study, we sought to address the potential vulnerabilities in biometric security, specifically focusing on capacitive fingerprint recognition. Our approach combined both our actual experimental procedures and a proposed attack scenario that could be realized in a broader context.

In our experimental setup, we initially focused on collecting real-life fingerprint samples from our research team and lab members, ensuring a variety of unique fingerprint patterns. Once the molds were set and ready, we poured fresh Alginate into them to create fingerprint replicas. These replicas were then used to attempt unauthorized access on a variety of IoT devices equipped with capacitive fingerprint sensors. Our objective was to assess the success rate of these replicas in deceiving the sensors and granting unauthorized access.

\begin{figure}[!ht]
  \centering
  \includegraphics[width=0.475\textwidth]{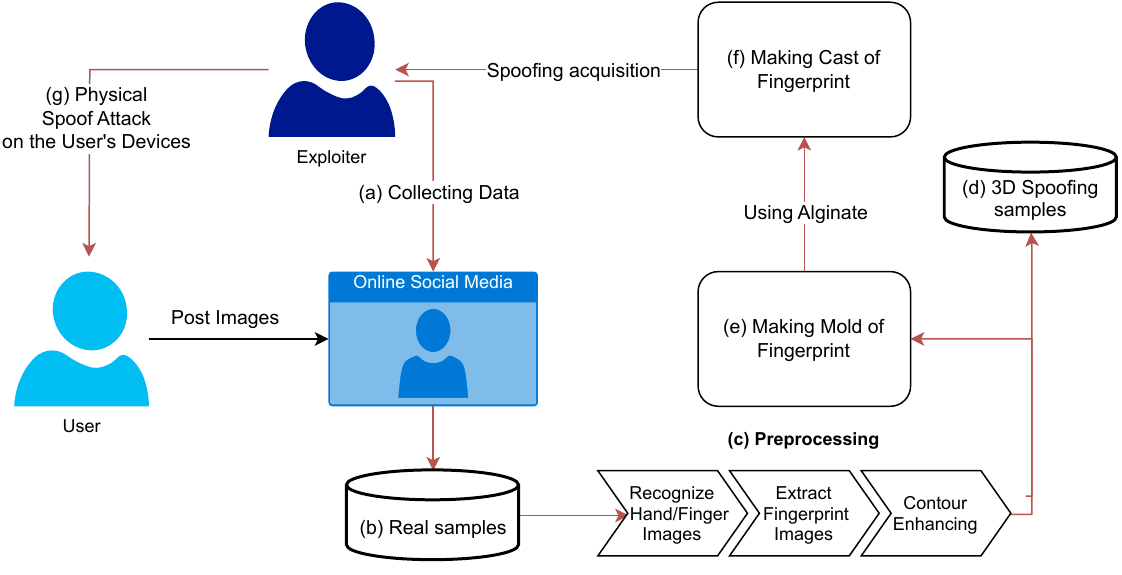}
  \caption{A potential scenario for spoofing attacks.}
  \label{figure2}
\end{figure}

Parallel to our hands-on experiment, we envisioned a more expansive attack scenario demonstrated in Figure~\ref{figure2} to provide a clearer understanding of our conceptualized approach. In this proposed scenario, adversaries would (a) collect high-resolution fingerprint images from online platforms, where users might inadvertently share their biometric data, and (b) store it. (c) Advanced image processing algorithms would then enhance these images, capturing every minute detail. The enhanced images would be transformed into (d) 3D models, which are stored as ready samples for spoof attacks and would then serve as blueprints for 3D printing. (e) The resultant 3D prints would act as molds for (f) creating fingerprint replicas, potentially using materials like Alginate. Finally, (g) exploiter will use the Alginate cast to conduct a physical spoof attack on the user's IoT device. This scenario functions as a detailed narrative of the fictitious attack process, providing readers with a comprehensive overview of the planned techniques and step which biometric security can be compromised in today's digital age. It is essential to stress that the scenario outlined here is merely a conceptual framework outlining the kinds of threats and attack techniques that an exploiter could use.

Together, our actual experiments and the proposed attack scenario highlight the pressing need for robust biometric security measures, especially as IoT devices become more integrated into our daily lives.

\subsection{Adversarial Model}
Understanding the potential adversary is paramount, and by modeling the capabilities, intentions, and methods of a potential attacker, we can better anticipate vulnerabilities and design systems to counteract these threats. In the context of our study, the adversarial model focuses on individuals or entities aiming to exploit the vulnerabilities in biometric authentication systems, particularly those using capacitive fingerprint recognition in IoT devices.

The adversary's operation can be broken down into distinct stages. Initially, they focus on data collection and processing. The adversary scours online platforms, such as social media sites and online marketplaces, to gather high-resolution images that might inadvertently display users' fingerprints. Once collected, these images undergo enhancement using advanced image processing algorithms, ensuring that even the minutest details of the fingerprint patterns are discernible. The enhanced images are then converted into 3D models.

Following this, the adversary shifts to physical replication and attack. With the 3D models at hand, the adversary employs precision 3D printing techniques, like DLP, to create molds of the fingerprints. Materials like Alginate, which closely mimics the texture and elasticity of human skin, are used to cast replicas from these molds. Armed with these replicas, the adversary attempts to deceive capacitive fingerprint sensors on IoT devices, seeking unauthorized access.

It's crucial to note that while our actual experiments were confined to using real-life fingerprint samples, the adversarial model we propose is broader, encompassing potential real-world scenarios. To validate our conceptual approach, future research should follow a structured methodology. Initially, this would involve the collection and analysis of high-resolution images from online platforms, focusing on those that might inadvertently display users' fingerprints. Advanced image processing techniques would be employed to enhance these images, ensuring that even the minutest details of the fingerprint patterns are discernible.

For the realization of our study, both in its actual implementation and the proposed concept, a combination of hardware and software tools was envisioned. In the actual experiments we conducted, our toolkit was straightforward yet effective. We utilized Petri dishes and mixing containers, which are essential for preparing and setting the Alginate. These simple tools allowed us to create molds and casts of real-life fingerprints, which were then tested on a specific fingerprint door lock available to our team.

\begin{table*}[h]
\centering
  \caption{Summary of Spoofing Attempts with Alginate}
\scalebox{0.64}{
\begin{tabular}{@{}lllll@{}}
\multicolumn{5}{c}{} \\
\toprule
\textbf{Device Type} & \textbf{Device Price}   & \textbf{Average Success Rate} & \textbf{Average Attempts of First successful outcome} & \textbf{Average global rating to number of reviews}\\
\toprule
IoT Smart Entry Door Lock 1         & \$140        & 3/20    & 7/20  &  0.0057       \\
IoT Smart Entry Door Lock 2         & \$250        & 20/20   & 1/20  &  0.0657       \\
IoT Smart Entry Door Lock 3         & \$90         & 0/20    & N/A   &  0.0005       \\
    \bottomrule
\end{tabular}}
 \label{tab:Table_2}
\end{table*}

In the broader concept we're proposing, a more extensive array of tools would be required. High-resolution images, whether from cameras, either professional-grade or advanced smartphone variants, or from images shared over platforms, would be pivotal for capturing detailed images of fingerprints from various surfaces and lighting. Transitioning from 2D to 3D, a Digital Light Processing (DLP) 3D printer would be recommended. Recognized for their precision, DLP printers can produce detailed prints, capturing the intricacies of fingerprints. Alginate, with its skin-like properties, is suggested as the primary casting material.

On the software spectrum, libraries like OpenCV would be essential for image processing, ensuring the fingerprint patterns retain their clarity and detail. After processing, 3D modeling tools such as Blender or Tinkercad would extrude these enhanced 2D images into 3D models. The YOLO system, known for its real-time object detection capabilities, would be instrumental in swiftly identifying hand and finger images from extensive datasets. Depending on the 3D printer in use, specialized software like Cura or PrusaSlicer might be needed to translate the 3D models into print-ready G-code formats.

Together, these tools, both from our actual experiments and the proposed concept, provide a comprehensive framework. This combination ensures a rigorous examination of biometric security measures under potential real-world attack scenarios. The outcomes of this line of research would have significant implications for both biometric security and the field of digital forensics. By understanding the vulnerabilities in current fingerprint recognition technologies and assessing the potential of new spoofing methods, we can pave the way for enhancing the security and reliability of biometric systems.

\subsection{Experiment Setup}
We carefully planned our experiment to ensure the accuracy, repeatability, and validity of the results. To simulate real-life situations where an exploiter might attempt unauthorized access using spoofed fingerprints, we conducted our experiments in a controlled laboratory environment. Particular attention was given to maintaining consistent temperature (Exactly 66$^{\circ}$F)  and humidity levels (Exactly 68\% relative humidity (RH)), as these factors significantly influence the setting time and consistency of Alginate.

For the actual experiments, we diverged from traditional fingerprint collection methods. Instead, we directly created molds from the fingers of three willing participants within our research team and lab members. This approach was chosen to ensure a direct and accurate impression of the fingerprints. Participants dipped their right thumb fingers into a specially prepared Alginate mixture, carefully formulated to ensure uniform texture and consistency. The meticulous submersion process for each participant's finger aimed to capture detailed imprints and avoid any distortions or air bubbles.

Quality control was paramount throughout the process. We established strict protocols for regularly checking and adjusting the Alginate mixture to maintain its optimal ratio and consistency. Each mold was thoroughly inspected for accuracy and completeness before being used to create Alginate casts. Using these direct molds, we then prepared Alginate casts to serve as the spoofed fingerprints. The casts were tested on three different IoT smart door locks equipped with capacitive fingerprint sensors, chosen for their widespread use in modern homes and offices. The selection of devices represented a range of price points and were popular consumer and marketplace choices, each manufactured within the past three years.

During the testing phase, each attempt to unlock the devices using the Alginate casts was meticulously recorded. In total, 180 separate tests were performed (20 attempts per device, across 3 devices, with 3 participants), providing a robust dataset for analysis. We noted the number of attempts made, the time taken for each attempt, and the outcomes (success or failure) of each trial. This data was then used to assess the success rate of the Alginate casts in spoofing the fingerprint sensors.

While valuable insights were gained into the efficacy of Alginate for spoofing attacks, our study did not directly compare the performance of these Alginate casts, other spoofing materials, and the actual subjects' fingerprints on the IoT devices. Such a comparison could have provided additional insights into the relative effectiveness of Alginate in mimicking actual fingerprints under practical conditions. For that, we will address our roadmap for future research later in the paper to benefit from incorporating this comparative analysis, offering a more holistic understanding of Alginate's potential to circumvent biometric security measures.

\section{Results} 
\label{sec:findings}
As previously discussed, our primary objective was to assess the efficacy of Alginate as a spoofing material for capacitive fingerprint sensors commonly found in IoT devices. Our experiments using Alginate casts to spoof fingerprint sensors were successful on various Home Lock IoT devices and have yielded noteworthy findings. The results indicated in the Table~\ref{tab:Table_2} shows Fascinating observations.

IoT Smart Entry Door Lock 1, situated at a mid range price point, demonstrated a moderate vulnerability to Alginate-based spoofing. The variability in success rates across different fingerprint molds suggests that the effectiveness of the spoofing technique might be influenced by the quality of the molds or the inherent characteristics of the device's sensor.

Surprisingly, IoT Smart Entry Door Lock 2, a higher-priced model, was consistently more vulnerable to spoofing attempts. This was evident from the fact that, on average, it took fewer attempts to achieve the first successful spoof. While it's tempting to correlate higher prices with better security, our findings challenge this notion.

IoT Smart Entry Door Lock 3, the most affordable among the tested devices, resisted all spoofing attempts, further emphasizing that price does not solely dictate the robustness of biometric security. 

Furthermore, we determined two other variables to understand the penetrability and trustability of these IoT smart lock devices. The ``Average Attempts of First successful outcome'' metric is particularly interesting. A lower number in this metric indicates higher vulnerability, as it means fewer attempts were needed to deceive the sensor. Conversely, a higher number suggests better resilience against spoofing. This metric underscores the importance of rigorous testing and validation to ensure the security and reliability of biometric systems as they become more prevalent in everyday devices.

The other variable is ``Average global rating to number of reviews,'' which is based on the rating data we collected from marketplaces that these devices are available on and the amount of people who reviewed these devices. As a 4.5 rating received from 400 buyers is not equal to the one received from 8000 people, we decided to standardize these ratings by dividing these two metrics to reach a number that shows the trustability of the device. A lower number in this metric indicates higher trustability of the device, and it positively correlates with the results we achieved. The lowest price device we got had the best performance against our spoofing attack and the highest rating with the most people rating and reviewing the device. On the other hand, the most expensive device we had was one of that marketplace's ``Choice'' devices, and it did not have the worst rating. However, the number of reviewers it had was much smaller than the other device, resulting in a higher metric figure and a weaker performance against spoofing in our tests.

Our study's success in spoofing various home lock IoT devices using Alginate casts has crucial implications for biometric security. The ability of Alginate, a natural and easily moldable material, to replicate fingerprints and deceive sensors underscores a significant vulnerability in widely-used biometric systems. This finding suggests that even basic materials can pose a threat to sophisticated security measures.

Furthermore, our experiments contribute to a broader understanding of biometric security's current limitations and call for continuous innovation in biometric technology, especially in developing anti-spoofing measures. The need for more robust authentication methods is highlighted by the effectiveness of Alginate in spoofing fingerprint sensors.

Additionally, the implications of our study extend into the domain of digital forensics. The ability to create fingerprint replicas challenges traditional forensic methods that rely heavily on fingerprint evidence. This necessitates new forensic techniques to differentiate between authentic and fabricated biometric data, ensuring the integrity of forensic investigations in an era where digital and biometric spoofing is increasingly plausible.

Our preliminary findings validate the hypothesis that Alginate can serve as an effective material for spoofing fingerprint sensors. The results not only underscore the potential vulnerabilities in current biometric security systems but also pave the way for a deeper exploration into the domain of digital forensics. The proposed concept of a novel attack vector targeting Smart Lock devices equipped with fingerprint sensors emphasizes the pressing need for robust countermeasures.

\section{Discussions} 
\label{sec:discussion}
The results of our study highlights the potential vulnerabilities inherent in capacitive fingerprint sensors, especially when confronted with sophisticated spoofing techniques using materials like Alginate. Our findings not only highlight the efficacy of Alginate as a spoofing material but also raise pertinent questions about the broader implications for IoT security.

One of the primary takeaways from our research is the realization that even as technology advances, the fundamental challenges associated with biometric authentication remain. While fingerprint sensors have become more sophisticated over the years, they are not infallible. The fact that a natural material like Alginate, which is readily available and easy to work with, can deceive these sensors is a testament to the persistent challenges in the realm of biometric security.

In the age of social media, where influencers frequently share personal moments online, a potential risk emerges with the advanced biometric security breach method using Alginate-based fingerprint replicas. Envision a scenario where a malicious actor, through meticulous analysis of an influencer's Instagram feed, discovers their residence and gathers high-resolution images showcasing their fingerprints. Utilizing these images, the adversary employs 3D printing technology to create precise fingerprint replicas from Alginate. This technique could allow unauthorized access to the influencer's home, exploiting biometric security measures to protect their privacy and safety. Implementing stricter privacy settings and advocating for social media platforms to develop algorithms that alert users to potential biometric data exposures can significantly mitigate risks. This speculative situation underscores the urgent need for reinforced security protocols in an increasingly interconnected and publicly exposed society.

The connection of our research to digital forensics and incident response is significant. Our findings necessitate a reconsideration of digital forensic practices, particularly in scenarios where spoofed fingerprints could be used to gain unauthorized access to devices containing critical evidence. The need for advanced detection techniques and new forensic methodologies to distinguish genuine from fabricated fingerprints is more crucial than ever. This research not only demonstrates a security vulnerability but also calls for collaborative efforts between cybersecurity and digital forensics communities to devise effective strategies for preventing, detecting, and responding to advanced spoofing attacks.

Our investigation into IoT device biometric security reveals a complex landscape where device price does not directly correlate with security efficacy. Intriguingly, the most budget-friendly device tested, the IoT Smart Entry Door Lock 3, was impervious to all spoofing attempts, challenging the assumption that higher cost ensures better protection. This finding underscores the importance of implementing the underlying technology and security features over price. Policymakers and regulatory bodies must conduct comprehensive evaluations of these devices and sensors, establishing institutions dedicated to testing and disseminating reliability scores for each sensor prior to getting released to the market and industry. Establishing a comprehensive framework for regular, updated assessments of security measures, facilitated by a coalition of manufacturers, cybersecurity experts, and policymakers, can ensure that security standards evolve alongside technological advancements. This public-private partnership model would encourage sharing best practices and contribute to developing universally robust biometric security protocols, reinforcing the importance of technology and implementation over price. Additionally, our analysis highlighted a nuanced relationship between device price, the average attempts required for a successful spoofing outcome, and the devices' global ratings concerning the number of reviews. These correlations that relate to the vulnerability and trustability of these capacitative sensors further emphasize the need for a detailed examination of biometric security measures, advocating for a shift in focus towards the efficacy of security implementations rather than mere cost.

Our research is meant to prompt a rethinking of how biometric security is implemented in IoT security devices while introducing a threat concept that we set to explore more around it in the future. As we progress toward a more interdependent global community, the stakes have never been higher. It's crucial that we keep ahead of would-be exploiters by constantly developing and improving our security protocols, especially with the emerging and rapid growth of artificial intelligence tools and image enhancement algorithms. Our findings point to potential weak spots, suggest new potential dangers, and pave the way for improved defenses on the road.

\section{Challenges and Limitations}
\label{sec:limitations}
This study has a number of limitations. Achieving a consistent mixture of Alginate was pivotal for the reliability of our tests. Even slight variations in the consistency could affect the sensor's ability to read the spoofed fingerprint accurately. Ensuring this uniformity in the Alginate mixture for every cast posed a significant challenge.

Another factor that introduced variability was the fingerprint sensors themselves. Different IoT devices, sometimes even those produced by the same manufacturer, might have sensors with distinct sensitivities and recognition algorithms. This inherent variability could lead to inconsistent results when testing across multiple devices.

While our research provides useful information about the security of fingerprint sensors in IoT devices, readers should proceed with care. While the IoT Lock Device 2 we utilized in our experiments did not display any glaring flaws, we cannot rule out the possibility of device-specific vulnerabilities. To get a full picture of the device's security, it would be helpful to do research on numerous devices of the same model to find whether it was this specific device that was weak towards our spoofing attack or if it is the same with all devices by this specific manufacturer.

In addition, the sample size of our study was small. This study only looked at the possibility of employing Alginate as a spoofing material; future research on the threat we proposed would explore a wider range of devices and materials to strengthen our findings and establish more certain conclusions. As a result, not only would our current findings be confirmed, but perhaps more nuances relating to biometric security in IoT devices would be uncovered as well.

Our controlled lab environment, while meticulously set up, might not mirror all real-world conditions. External factors such as ambient temperature, humidity, and the natural wear-and-tear of the sensor can influence the success rate of a spoofing attack.

For the proposed concept attack, the quality of the sourced fingerprint images is of utmost importance. Low-resolution images or those with obscured fingerprints can diminish the accuracy of the 3D-printed molds, thereby affecting the overall success rate of the attack.

Finally, while 3D printing is advanced, it has its own set of limits. The precision required to capture the minute minutiae of a fingerprint pattern is immense, and even little differences in printer accuracy or the printing material (e.g. resin or filaments) might result in recognition failures.

\section{Conclusions and Future Work} 
\label{sec:conclusion}
The use of capacitive fingerprint recognition technology within the rapidly expanding domain of IoT devices has introduced a range of advantages and complexities. Although the authentication procedures mentioned above provide convenient access, our research has revealed their vulnerability to sophisticated spoofing techniques, specifically through the use of Alginate. The potential of Alginate to circumvent the biometric security of IoT devices is underscored by its deceiving qualities, as well as its inherent features that closely resemble the conductivity and resistance of human skin.

Our proof-of-concept, which involved crafting molds directly from genuine fingerprints and subsequently replicating them using Alginate, showcased the material's remarkable ability to mimic real fingerprints both visually and tactilely. This study raises concerns about potential breaches of individual privacy, especially in light of the novel image identification algorithms and attack scenarios we provided. The concept of extracting fingerprint images from publicly available sources, enhancing them, and then creating 3D models for spoofing purposes remains a significant area of concern and future exploration.

While our experiments provide foundational insights, they do not encompass the full scope of the proposed novel vulnerability involving public image sources. The exploration of these scenarios underscores the necessity for increased awareness and the development of preventive measures. This includes educating the public about the risks associated with sharing high-resolution images and urging social media platforms to implement technologies that obscure fingerprint details. Additionally, it highlights the imperative for more advanced anti-spoofing technologies in biometric systems to counteract these potential threats. This speculative examination provides a foundation for future research, emphasizing the need for ongoing innovation in the field of biometric security and digital forensics to safeguard against the evolving methods of biometric data exploitation. Further research is required to explore this aspect comprehensively.

Our future research efforts will focus on these aspects, aiming to validate the theoretical model we have proposed. By addressing these challenges, we intend to provide a more comprehensive understanding of the risks associated with fingerprint data exploitation, particularly in the context of IoT device security. This exploration will not only contribute to the technological understanding of biometric spoofing but also emphasize the critical need for enhanced security measures in the rapidly evolving digital landscape. As we look to the future, the structured hypotheses we formulated—both device-specific and material-specific will serve as foundational pillars for subsequent investigations. These hypotheses not only encapsulate our current understanding but also set the stage for a series of controlled experiments. Specifically, we aim to rigorously evaluate the interplay between different materials and device models in the context of spoofing attacks, with a keen focus on validating or refuting our initial postulations about Alginate's efficacy and the vulnerabilities of specific IoT devices.

Our research makes it evident that the digital forensics community needs to step up its efforts to develop effective anti-spoofing solutions. As IoT Smart locks become more commonplace, protecting our private information becomes more than simply a technological difficulty; it's a social necessity. Our findings reveal new information and sound the alarm for increased attention and new approaches to security as the IoT infrastructure continues to develop at an exponential rate.

\section{Acknowledgment}
The authors would like to thank Seth Barrett, Rajon Bardhan and Bradley Boswell, Graduate Research Assistants in the School of Computer \& Cyber Science, Augusta University, for their support and motivation.



\bibliographystyle{agsm}
\bibliography{main.bib}
%

\end{document}